# Phase separation and surface segregation in ceria-zirconia solid solutions

Ricardo Grau-Crespo,[1]* Nora H. de Leeuw,[1] Said Hamad[2] and Umesh V. Waghmare[3]

[1] *Department of Chemistry, University College London, 20 Gordon Street, London WC1H 0AJ, UK.*

[2] *Instituto de Ciencia de Materiales de Sevilla, CSIC–Universidad de Sevilla, Avda. Américo Vespucio 49, 41092 Sevilla, and Department of Physical, Chemical and Natural Systems, University Pablo de Olavide, Carretera de Utrera, km 1, 41013 Sevilla, Spain.*

[3] *Theoretical Sciences Unit, Jawaharlal Nehru Centre for Advanced Scientific Research, Jakkur Campus, Bangalore 560 064, India.*

*Author for correspondence (r.grau-crespo@ucl.ac.uk)

Using a combination of density functional theory calculations and statistical mechanics, we show that a wide range of intermediate compositions of ceria – zirconia solid solutions are thermodynamically metastable with respect to phase separation into Ce-rich and Zr-rich oxides. We estimate that the maximum equilibrium concentration of Zr in $CeO_2$ at 1373 K is ~2%, and therefore equilibrated samples with higher Zr content are expected to exhibit heterogeneity at the atomic scale. We also demonstrate that in the vicinity of the (111) surface, cation redistribution at high temperatures will occur with significant Ce enrichment of the surface, which we attribute to the more covalent character of Zr-O bonds compared to Ce-O bonds. Although the kinetic barriers for cation diffusion normally prevent the decomposition/segregation of ceria-zirconia solid solutions in typical catalytic applications, the separation behaviour described here can be expected to occur in modern three-way catalytic converters, where very high temperatures are reached.

## 1. Introduction

Three-way catalytic converters (TWC) are devices employed for the reduction of harmful emissions from car exhausts, by catalysing three simultaneous processes: (i) the reduction of nitrogen oxides ($NO_x$) to nitrogen and oxygen, (ii) the oxidation of carbon monoxide to carbon dioxide, and (iii) the oxidation of unburnt hydrocarbons (HC) to carbon dioxide and water. The catalyst typically consists of a combination of noble metals, including Rh to promote $NO_x$ reduction and Pt or Pd for CO/HC conversion, a ceria-based material to promote the so-called oxygen storage capacity (OSC), and a high-surface-area $Al_2O_3$ support to achieve good catalyst dispersion. The incorporation of zirconium in the ceria component, forming the solid solution $Ce_{1-x}Zr_xO_2$, is known to improve the performance of the catalyst by enhancing the resistance of the material to sintering, and at the same time increasing the reducibility of the oxide, leading to more oxygen vacancies which are responsible for the activity of the catalyst towards oxygen-containing molecules (Di Monte & Kaspar 2005).

The so-called close-coupled converter is nowadays employed to improve efficiency of TWCs during cold start-up of the engine (Heck & Farrauto 2001). This converter is mounted directly on the engine exhaust manifold, which exposes the catalyst to temperatures as high as 1373 K. Therefore, the stability of the properties of ceria-zirconia solid solutions at these high temperatures is an issue of concern. For example, it has been previously reported that ceria-zirconia solid solutions experience phase separation into Ce-rich and Zr-rich oxides upon treatment at temperatures of 1273 – 1473 K (Yashima *et al.* 1994; Bozo *et al.* 2001; Di Monte *et al.* 2004). The effect of composition heterogeneities on the OSC of ceria-zirconia solid solutions are not as yet well understood, with some authors arguing that homogeneity improves the OSC (Nagai *et al.* 2002), while other authors have reported an increase in efficiency with nanometric-level heterogeneity (Mamontov *et al.* 2003).

It is clear that a systematic understanding of the properties of ceria-zirconia solid solutions at high temperatures is required in order to provide a rational basis for the design of more efficient TWC catalysts. This understanding is hard to achieve only from experimental investigation of the real catalysts, as their structural and functional complexity makes it difficult to separate the effects from different components. Computer simulations of model systems can therefore be of great utility. Initial simulations of ceria – zirconia solid solutions were performed more than a decade ago by Balducci *et al.* (1997); Balducci *et al.* (1998), based on classical interatomic potentials and a mean field approach to represent hybrid $Ce^{4+}/Zr^{4+}$ ions. Although this approximation allows a simple calculation of some effective structural parameters in the system (e.g. cell parameters), it cannot account for any effects related to the local distribution of the cations. Conesa (2003) has compared the stabilities of different cation distributions for the 50:50 ($Ce_{0.5}Zr_{0.5}O_2$) solution and found small but significant energy differences between configurations, which would be missed within a mean-field approach. He also concluded that, in comparison with density functional theory (DFT) calculations, interatomic potential simulations tend to exaggerate the lattice distortions and the energy differences between configurations in the solid solution. Based on a combination of DFT calculations and EXAFS measurements Dutta *et al.* (2006) attributed the origin of the enhanced OSC to the presence of a distribution of strongly and weakly bound oxygen atoms in the ceria-zirconia bulk, which is not present in pure ceria. The surface redox behaviour of the solid solution has been studied theoretically by Yang *et al.* (2008a; b), who found that Zr ions induce severe distortions in the surface structure, generating non-equivalent oxygen ions, with the lowest vacancy formation energies corresponding to the oxygen ions neighbouring the Zr dopant. These authors suggest that the smaller $Zr^{4+}$ dopants provide some spare space for accommodating the bigger $Ce^{3+}$ cations resulting from the formation of an oxygen vacancy.

In the present work, we employ density functional theory techniques to investigate the phase separation behaviour of ceria-zirconia solid solutions at high temperatures. We will focus on the Ce-rich part of the solid solution ($0<x<0.5$ in $Ce_{1-x}Zr_xO_2$), which exhibits cubic symmetry (Cabanas *et al.* 2001; Lee *et al.* 2008). The stability of the bulk solid solution with respect to the pure oxide phases and the segregation behaviour at the (111) surface are discussed, and our results are compared with available experimental information.

## 2. Calculation details

The DFT calculations were performed with the VASP code (Kresse & Furthmuller 1996a; b), using the generalized gradient approximation (GGA), with a density functional built from the local functional of Perdew & Zunger (1981), and the gradient corrections by Perdew *et al.* (1992). In order to correct for the limitations of the DFT method in the description of the highly localised *f* orbitals, we have employed the so-called DFT+U method, which introduces an energy term that penalises the hybridisation of the Ce $f$ orbitals with the O ligands, thereby forcing the localisation of the $f$ electron (Fabris *et al.* 2005; Nolan *et al.* 2005b; Andersson *et al.* 2007). We use a Hubbard parameter $U_{eff}$ = 5 eV, which is the value typically used in studies of ceria surfaces (Nolan *et al.* 2005a; Wilson *et al.* 2008; Hernandez *et al.* 2009) and has also been recommended by Andersson *et al.* (2007) based on a systematic study of CeO$_x$ oxides.

The interaction between the valence electrons and the core was described with the projected augmented wave (PAW) method (Blochl 1994) in the implementation of Kresse & Joubert 1999. The core levels up to $4p$ in Zr, $4d$ in Ce and $1s$ in O were kept frozen in their atomic reference states. The number of plane waves in VASP is controlled by a cutoff energy, which was set to the default value of $E_{cut}$=400 eV in all our calculations except for the volume relaxations in the bulk, where an increased cutoff of 500 eV was employed. It was checked that these values provided well converged results for energies and geometries.

The Ce$_{1-x}$Zr$_x$O$_2$(111) surface is represented in our calculations by oxygen-terminated slabs, which repeat periodically in the direction perpendicular to the surface, separated by a vacuum gap of ~15 Å. Each slab has two symmetrically equivalent surfaces and contains 18 atomic layers (6 O-M-O tri-layers, where M=Ce or Zr). This is double the thickness that we have employed in previous simulations of cubic zirconia and ceria (111) surfaces (Grau-Crespo *et al.* 2007b; Branda *et al.* 2009; Grau-Crespo *et al.* 2009), as in this case we want to investigate the properties of different atomic layers and not only the behaviour of the top surface layer. Parallel to the surface, the supercell consists of a 2 × 2 array of hexagonal surface unit cells. Each unit cell contains one MO$_2$ unit at the surface, and therefore the simulation supercell has four oxygen ions at each surface, and overall composition M$_{24}$O$_{48}$ (Fig. 1). In modelling the bulk, a cell with similar shape and orientation of the axes was employed, except that there was no vacuum gap and the number of atomic layers in the $c$ direction was reduced to 9 (all layers are now symmetrically equivalent), leading to a M$_{12}$O$_{24}$ supercell. Each structure was fully relaxed to the equilibrium geometry using a conjugate gradients algorithm, which stops when the forces on the atoms are all less than 0.01 eV/Å. A 3×3×1 k-point mesh was used to sample the reciprocal space for Brillouin zone integrations.

## 3. Results and discussion

*(a) Mixing thermodynamics*

In order to investigate the properties of the solid solution at a given composition, we generate all the symmetrically inequivalent configurations with the same composition in the supercell. The total number of configurations with composition $Ce_{N-n}Zr_nO_{2N}$ is $N!/(N-n)!n!$ ($N=12$ is the number of formula units in the supercell). For example, for the 50:50 solid solution, the total number is 924. However, the computational load can be reduced considerably by calculating only those calculations which are symmetrically inequivalent, for example, only 17 for the 50:50 solution. Table 1 shows the total number and the reduced number of configurations for each composition, as determined using the the SOD program. (Grau-Crespo *et al.* 2007a). For each of these configurations, we obtained the degeneracy (number of times the configuration is repeated in the full configurational space) and the energy at the DFT level.

Our first observation is that, for all compositions, the lowest-energy configurations are those where all the Zr ions are grouped together, preferably forming clusters or full Zr layers in the *ab* plane. This separation indicates a tendency to ex-solution. Our model with periodic boundary conditions does not allow us to investigate the real extension of the heterogeneity, but we can quantify the tendency to ex-solution within bulk phases by calculating the enthalpy of mixing:

$$\Delta H_{mix} = H[Ce_{1-x}Zr_xO_2] - (1-x)H[CeO_2] - xH[c\text{-}ZrO_2] \qquad (1)$$

where $H[CeO_2]$ and $H[c\text{-}ZrO_2]$ are the DFT energies per formula unit of ceria and cubic zirconia, respectively, and $H[Ce_{1-x}Zr_xO_2]$ is the effective energy of the solid solution, calculated as an average of the DFT energies per formula unit of the different configurations at the given composition, weighted by a probability term that depends on the configuration energy $E_m$, its degeneracy $\Omega_m$ and the temperature $T$ (Grau-Crespo *et al.* 2000; Benny *et al.* 2009; Smith *et al.* 2010):

$$H[Ce_{1-x}Zr_xO_2] = \frac{\sum_m E_m \Omega_m \exp(-E_m/RT)}{N \sum_m \Omega_m \exp(-E_m/RT)} \qquad (2)$$

where $R$=8.49 J/molK is the gas constant. The resulting enthalpy of mixing is strongly positive, in agreement with recent calorimetric measurement by Lee *et al.* (2008) (Figure 2). We find that the result is practically independent on temperature for $T$>1000 K and therefore we have only plotted the enthalpies of mixing in the limit of full disorder ($T\rightarrow\infty$). Assuming a regular solid solution model (*e.g.* Prieto 2009, Ruiz-Hernandez *et al.* 2010), we can try to fit the data with a polynomial of the form:

$$\Delta H_{mix} = Wx(1-x) \qquad (3)$$

as in previous experimental work (Du *et al.* 1994; Lee *et al.* 2008). In principle, this fitting is not adequate in our case because the polynomial (3) implies a maximum at $x$=1/2, while our calculated enthalpies of mixing show a maximum at $x\approx$1/3. However, the slight decrease of the mixing enthalpy for 1/3<$x$<1/2 is likely to be an artefact resulting from the simulation cell size. At these intermediate compositions, some ordered structures appear that are somewhat lower in energy, in agreement with previous theoretical work (Conesa 2003), and their weight in the average (2) is exaggerated within the relatively small configurational space corresponding to our simulation cell. A proper investigation of the cation distribution at intermediate compositions would require a larger simulation cell. Therefore, we have excluded the last two points from the fitting with the polynomial form (3), which gives $W$=38 kJ/mol. This result is intermediate between the value of 28 kJ/mol obtained by Du *et al.* (1994) from fitting a regular solution model to experimental solubility data, and the value of 51 kJ/mol obtained by Lee *et al.* (2008) by fitting directly to calorimetric measurements.

The positive values of the enthalpy of mixing suggest that cation ordering is not a stabilizing factor in ceria-zirconia solid solutions, at least for the compositions examined here, and confirm that the Zr ions have an energetic preference to segregate or form a separate Zr-rich phase. Real samples, however, where homogeneity at the atomic level has been achieved using special synthesis methods (*e.g.* Cabanas *et al.* 2000; 2001), might not experience this trend unless subjected to temperatures high enough to overcome the cation diffusion barriers. The origin of this difficulty to mix is the difference between the ionic radii of the cations ($r$[Ce$^{4+}$]=0.97 Å and $r$[Zr$^{4+}$]=0.84 Å, for 8-fold coordination, according to Shannon 1976). The behaviour of the CeO$_2$-ZrO$_2$ contrasts with that of another well known fluorite-type solid solution, ZrO$_2$-Y$_2$O$_3$, where short-range-ordered domains appear to exist (Goff *et al.* 1999), consistent with a negative interaction parameter $W$ (Lee *et al.* 2003).

The discussion above is based on the energetics of mixing, but in order to describe the behaviour of the solid solution at any finite temperature we should also calculate entropies and free energies of mixing. It can be checked that, even assuming ideal (maximum) configurational entropy:

$$S_{\text{ideal}}(x) = -R[x \ln x + (1-x) \ln(1-x)] \qquad (4)$$

the resulting free energy of mixing $\Delta G_{\text{mix}} = \Delta H_{\text{mix}} - TS_{\text{ideal}}$ is positive except for very small values of *x*. Furthermore, since Zr-rich phases are known to be monoclinic (Garvie 1970) at the temperatures of interest here, the mixing free energy should be calculated with respect to the more stable monoclinic zirconia phase (m-$ZrO_2$), which makes the mixed phase even less stable with respect to phase separation. In order to estimate the solubility limit of Zr in $CeO_2$ we now consider the mixing free energy function:

$$\Delta G_{\text{mix}}(x,T) = Wx(1-x) + \Delta H_t x + RT[x \ln x + (1-x) \ln(1-x)] \qquad (5)$$

where we have introduced the enthalpy of the monoclinic-cubic zirconia phase transformation $\Delta H_t$ =8.8 kJ/mol (Navrotsky *et al.* 2005). The use of the ideal entropy is justified because we are interested in the region of very low Zr content, where the disorder should be nearly perfect. This analytical function allows us to interpolate our results to *x* values smaller than those directly obtainable with our simulation supercell, and its minimum with respect to *x* at a given temperature provides an estimation of the solubility limit.

Figure 3 shows that the maximum equilibrium solubility of Zr from monoclinic zirconia into the ceria structure is ~0.4 mol% at 973 K, and increases to 2 mol% at 1373 K. Thus, although ceria-zirconia solid solutions in the whole range of compositions can be synthesised under adequate conditions (*e.g.* Cabanas *et al.* 2000), our results taken together with previous experimental evidence clearly show that these solid solutions are metastable with respect to phase separation into Ce-rich and Zr-rich phases. This phase separation can actually occur in a close-coupled catalytic converter, where temperatures of up to 1373 K could lead to rearrangement of the cations in the solid solution.

We have also calculated the variation of the cubic cell parameter of the metastable solid with respect to composition (Figure 4). The small fluctuations in the cell parameters among different configurations for a given composition, which in some cases slightly break the cubic symmetry, were averaged out by calculating the mean volume of the supercell ($<V_{sc}>$) and from there the effective cell parameter as $a=(<V_{sc}>/3)^{1/3}$ (the division by 3 accounts for the different in volume between our supercell and the cubic unit cell). The theoretical result corresponds to the configurational average in the high-temperature limit, where configurations are only weighted by their degeneracies. The rate of the variation is linear, following the so-called Vegard's law (Vegard 1921), with the cubic parameter $a$ decreasing with Zr content as expected from the smaller radius of the $Zr^{4+}$ cation with respect to $Ce^{4+}$ and in agreement with experimental measurements (Cabanas *et al.* 2001; Lee *et al.* 2008). The calculated variation rate d$a$/d$x$=-0.31 Å is practically identical to the experimental rate, although the cell size is systematically overestimated in the DFT calculations by ~0.08 Å (1.5%).

*(b) Cation segregation at the (111) surface*

The surface calculations were performed by fixing the cell parameters of the slab to those of the bulk at the same composition. The number of configurations in the slab is drastically reduced by only including those keeping the inversion symmetry of the cell and then selecting the symmetrically inequivalent ones (Table 1). Since we are requiring that each side of the slab is identical to the other side by the inversion symmetry, the total number of configurations in each half slab, with composition $Ce_{12-n}Zr_nO_{12}$, is the same as in the bulk cell employed in the previous section. However, the cation layers perpendicular to the (111) surface are no longer equivalent to each other in the slab model, and therefore the number of symmetrically inequivalent configurations for the surface calculations is somewhat higher than for the bulk calculations.

The equilibrium zirconium content of a particular cation layer parallel to the (111) surface, depends both on the overall zirconium content of the slab and on the temperature, and can be calculated by taking the average:

$$c_l = \frac{\sum_m f_{ml} \Omega_m \exp(-E_m/RT)}{\sum_m \Omega_m \exp(-E_m/RT)}$$

where $f_{ml}$ is the fraction of sites occupied by Zr in the layer $l$ for configuration $m$. The results are shown in Fig. 5 for temperatures between 800 and 1600 K.

The most obvious feature of the cation distribution is the low concentration of Zr at the top (111) layer. Even for the 50:50 solid solution, at the highest temperature considered (1600 K), the equilibrium Zr content of the surface is only ~10%. The dependence of the calculated concentrations on temperature is relatively weak, especially at the top layer, but it is clear that increasing temperatures lead to more homogeneity in the composition of the interior of the slab, by equalising the Zr content in the second and third layers. Therefore, the segregation behaviour can be described as top-layer avoidance of Zr ions. This is consistent with the higher (111) surface energy of pure cubic zirconia (1.08 J/m$^2$) compared to pure ceria (0.74 J/m$^2$). Our calculated surface energies for the pure oxides are lower but exhibit the same trend as those calculated by Gennard *et al.* (1999) using Hartree-Fock calculations (1.48 J/m$^2$ for cubic zirconia and 1.24 J/m$^2$ for ceria). We note that the same trend in surface energies was found by these authors for the (011) surface (2.41 J/m$^2$ for cubic zirconia and 2.11 J/m$^2$ for ceria), which suggests that this surface will experience a similar segregation behaviour. As pointed out by Gennard *et al.*, this trend can be explained in terms of the more covalent character of the metal-oxygen interaction for Zr than for Ce. Since the (111) surface termination involves the cutting of some metal-oxygen bonds, the unsaturated coordination should be more easily accommodated by Ce cations than by the more covalent Zr cations.

Thus, according to our results, the redistribution of cations at high temperatures should occur with significant Ce-enrichment of the (111) surface of ceria-zirconia, regardless of the overall composition of the solid solution. Considering that this surface is the most prominent one in these fluorite oxides, and that other surfaces are likely to exhibit similar behaviour, we can expect the segregation effect to be noticeable in experimental measurements. Indeed, several experimental studies have presented comparisons of the bulk and surface compositions in ceria-zirconia samples, but the results are somewhat contradictory. For example, Daturi *et al.* (1999), using both X-ray photoelectron spectroscopy (XPS) and the band intensities of adsorbed methoxy species, found no evidence of surface segregation in ceria-zirconia samples in the full range of compositions, while Bozo *et al.* (2001), using XPS and ion scattering spectroscopy, found that phase separation at high temperatures occurred with Zr-enrichment at the surface. However, a number of other experimental studies have found Ce enrichment at the surface of the solid solution, in agreement with our theoretical prediction. Sun & Sermon (1996) measured a surface Ce/Zr ratio of 0.25 in a sample with nominal bulk composition Ce/Zr=0.15, and Martinez-Arias *et al.* (2003) obtained a Ce/Zr surface ratio of 2.6 for a sample with nominal bulk composition Ce/Zr=1. More recently, Ce-enrichment of the surface has been reported by Damyanova *et al.* (2008) for Zr-rich solutions (Ce/Zr=0.047 at the surface in a sample with only Ce/Zr=0.007 overall) and Atribak *et al.* (2009) for Ce-rich solutions (Ce/Zr=8.1 at the surface, when nominal composition was Ce/Zr=5.7), in both cases based on XPS measurements.

We can expect that the extent to which Ce-enrichment is observed in a specific sample depends on the thermal history of the sample. For example, the samples in the work by Daturi *et al.* (1999) were treated with temperatures (873 K) that were not high enough to allow cation diffusion and redistribution, which possibly explains the absence of cation segregation in their samples. The observation of a Zr-rich surface by Bozo *et al.* (2001) is more intriguing, and suggests that other factors, including kinetic ones, or opposite segregation behaviour of other surfaces in the solution, could be present, which requires further investigation. It is also important to realise that the XPS technique, which is employed to investigate the surface composition in most of the studies cited above, is sensitive not only to the top surface layer of the solid (where we predict the segregation effect to be most marked) but extends several layers into the bulk, and therefore direct comparison of XPS results with our theoretical predictions should be treated with caution.

## 4. Conclusions

Our simulation study has shown that ceria-zirconia solid solutions are metastable with respect to Ce-rich and Zr-rich oxides. We predict a maximum equilibrium solubility of Zr (from monoclinic zirconia) in $CeO_2$ of 2% at 1373 K. Therefore, we can expect some level of phase separation to occur in the solid solution when it experiences the high temperatures of a close-coupled catalytic converter, especially after long using times.  We have also shown that segregation of Zr will occur in the solid solution with Ce enrichment of the surface, at least in the most prominent (111) surface, where we argue that this behaviour is consistent with the more covalent character of Zr-O bonds compared to Ce-O interactions. Since the enhancement of the oxygen storage capacity provided by ceria-zirconia appears to be linked to the presence of both Ce and Zr ions around the oxygen vacancies (Dutta *et al.* 2006; Yang *et al.* 2008a; b), the separation/segregation behaviour described here is likely to have important consequences for the performance of Ce-Zr oxides as components of modern three-way catalysts, and therefore merits further investigation.

RGC's visit to JNCASR was supported by an international travel grant from the Royal Society. This work was funded by EPSRC grant EP/C51744X, and made use of the facilities of HPCx and HECToR, the UK's national high-performance computing services, via our membership of the UK's HPC Materials Chemistry Consortium, which is funded by EPSRC grant EP/F067496. We thank Prof. A. Navrotsky for making a manuscript available to us prior to publication and Dr J. A. Darr for useful discussions.

**Table 1.**

| $x$ in $Ce_{1-x}Zr_xO_2$ | $n$ in $Ce_{12-n}Zr_nO_{24}$ | Total number of configurations | Number of inequivalent configurations | |
|---|---|---|---|---|
| | | | bulk | surface |
| 0.083 | 1 | 12 | 1 | 3 |
| 0.167 | 2 | 66 | 3 | 9 |
| 0.250 | 3 | 220 | 6 | 20 |
| 0.333 | 4 | 495 | 11 | 42 |
| 0.417 | 5 | 792 | 13 | 57 |
| 0.500 | 6 | 924 | 17 | 66 |

**Figure captions**

**Figure 1.** Simulation cell employed to model the (111) surface of the ceria-zirconia solid solution. White balls are Ce or Zr cations, and darker (red in colour version) balls are oxygen anions. Dark lines show the 2x2 extension of the supercell. The images of the atoms at the cell border are included for clarity.

**Figure 2.** Calculated enthalpies of mixing in comparison with experimental results of Lee *et al.* 2008). The curved line represents the fitting of a regular-solution quadratic polynomial to the calculated values for low Zr concentrations.

**Figure 3.** Free energies of mixing for low Zr concentrations, as obtained from Eq. 5. The vertical dotted lines mark the solubility limit of Zr in $CeO_2$ at the particular temperature.

**Figure 4.** Calculated variation of the cubic cell parameter in comparison with the experimental results of Cabanas *et al.* (2001) and Lee *et al.* (2008).

**Figure 5.** Calculated equilibrium concentrations of Zr as a function of the distance to the (111) surface. Because of the slab construction, layers 1, 2 and 3 are equivalent to layers 6, 5 and 4, respectively.

**Figure 1.**

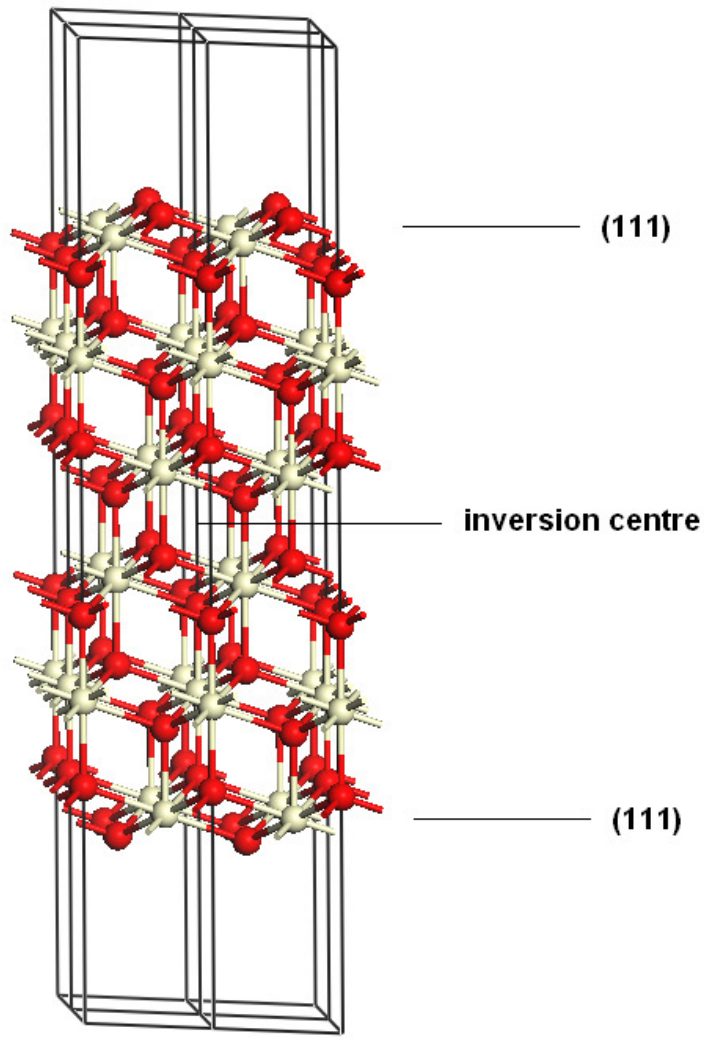

**Figure 2**

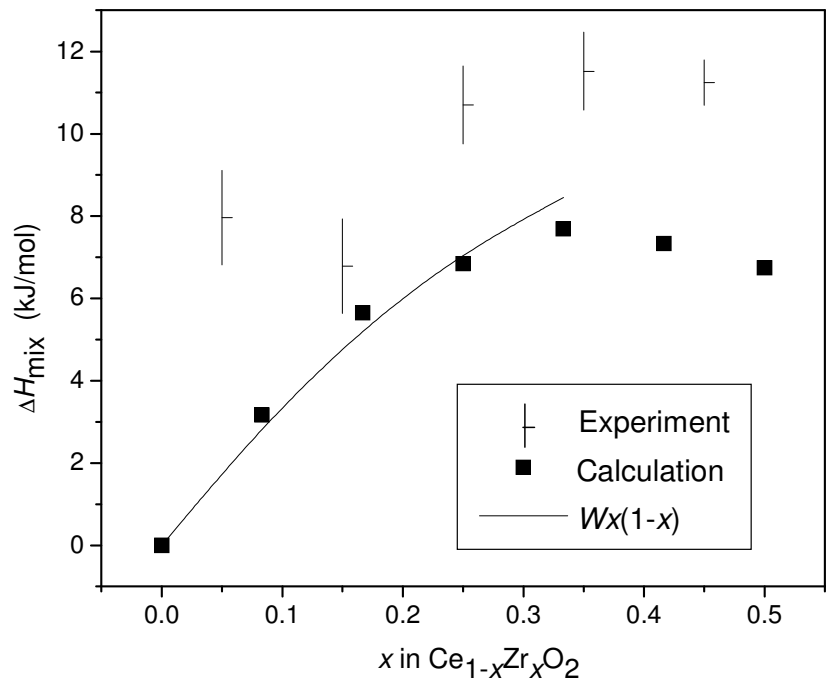

**Figure 3**

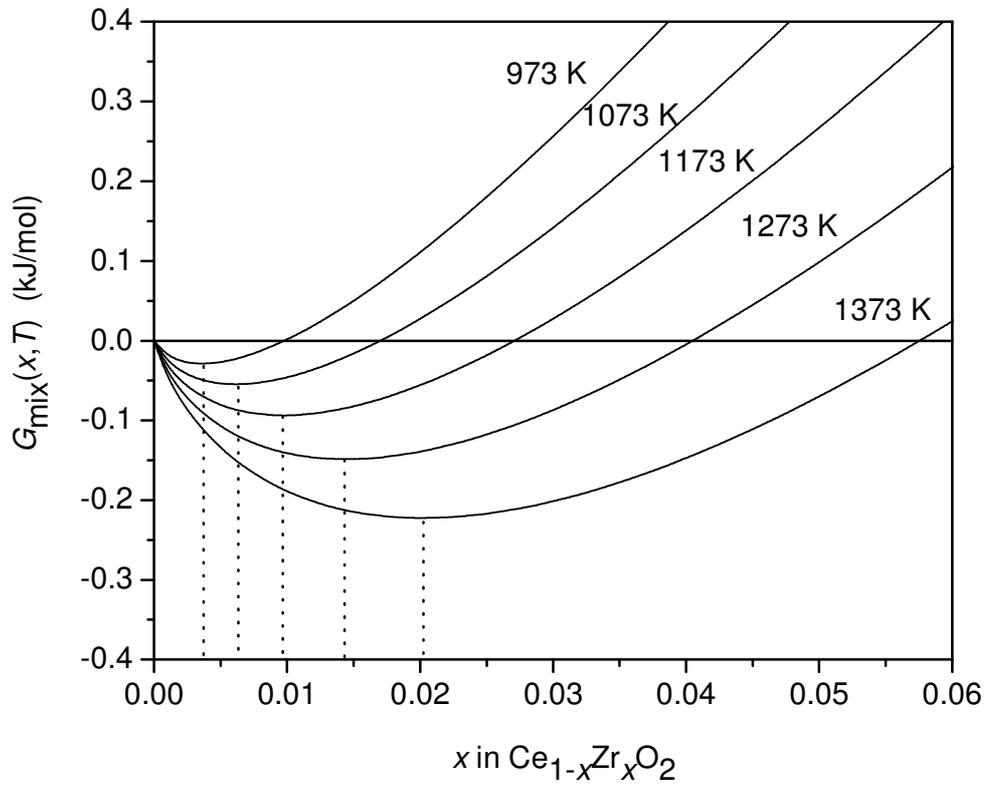

**Figure 4**

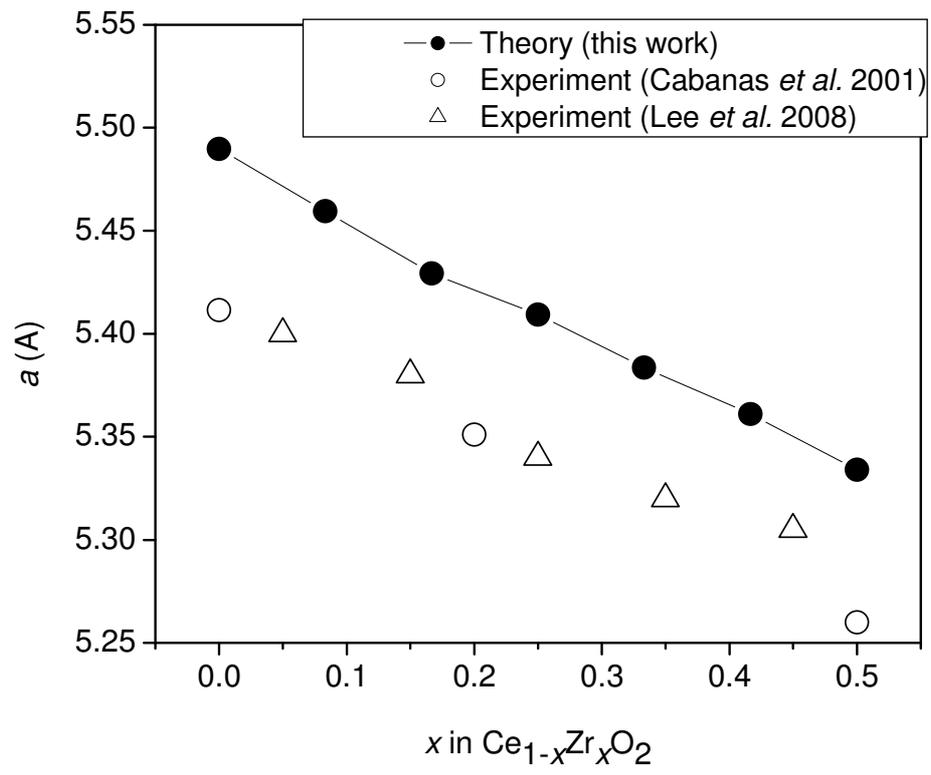

**Figure 5**

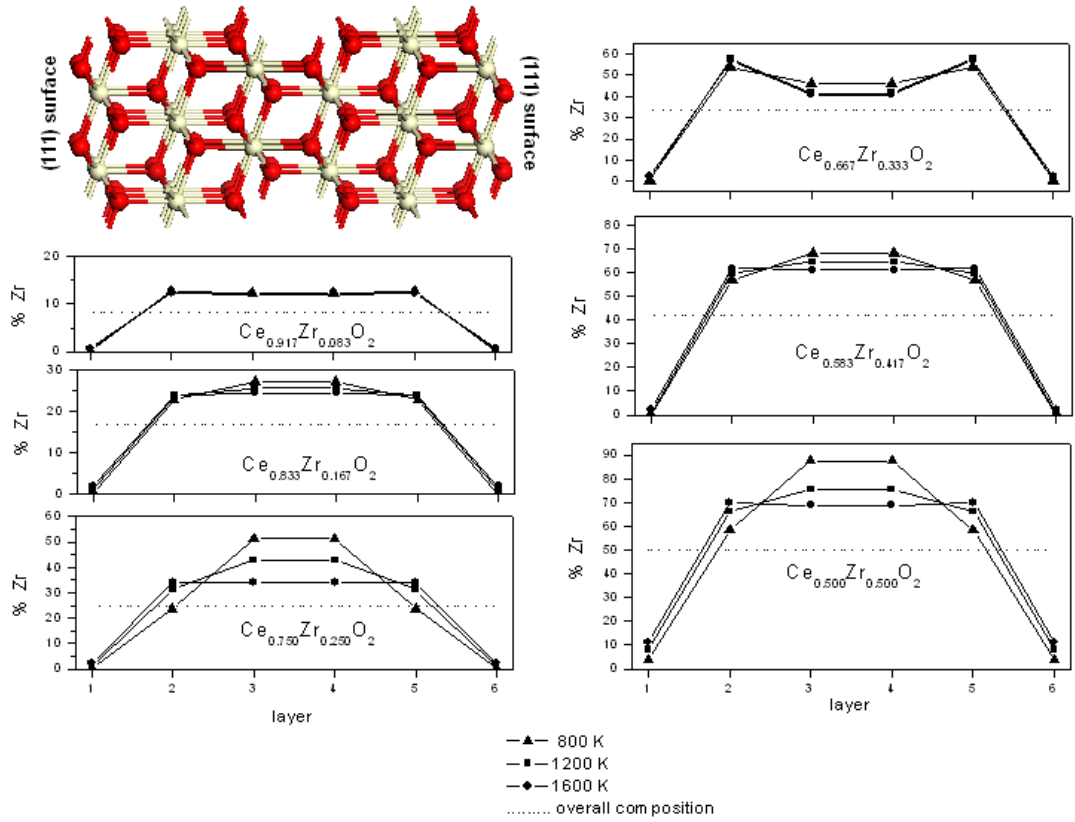